# 13.56MHz Rectifying Diodes Based on Metal Halide Perovskite

Peng Jin[1], Xuehui Xu[1], Zeng Chen[1,2], Xu Chen[1], Tianyu Liu[1], Hanbo Zhu[1], Xinya Chen[1] & Yang (Michael) Yang[1]*

[1]State Key Laboratory of Modern Optical Instrumentation, College of Optical Science and Engineering, Zhejiang University, Hangzhou, Zhejiang, China, 310027.

[2]Center for Chemistry of High-Performance & Novel Materials, Department of Chemistry, Zhejiang University, Hangzhou, Zhejiang, China, 310027.

**Corresponding Author:**

Email:yangyang15@zju.edu.cn[1]

**Abstract:** The increasing use of portable and wireless technologies has led to a growing focus on radio-frequency identification (RFID) tags. Among the various devices in RFID tags, rectifying diodes are the most demanding in terms of high-frequency performance, and these diodes are dominated by organic materials. However, their intrinsic low carrier mobility largely limits the rectifying ability of organic diodes. As an alternative, metal halide perovskites (MHPs) possess high carrier mobility, making them potential candidates for higher-frequency applications. Whereas their ion-migration issue may deteriorate their high-frequency performance. In this study, we report rectifying diodes based on MHPs that can rectify an incoming sinusoidal signal at 13.56 MHz. The diodes exhibit a high rectification ratio of $1.9 \times 10^3$ at ± 1 V and can rectify signals at even higher frequencies. We designed a triangular wave detection method to measure the intensity of ion-migration at different frequencies. Interestingly, the ion-migration did not occur at such a high frequency. The high-frequency stagnant ions and excellent carrier mobility make MHPs unexpectedly suitable for high-frequency applications, providing a promising solution to ion-migration issues and paving the way for perovskites in high-frequency areas.

With the increasing demand for wireless automatic identification procedures, radio-frequency identification (RFID) tags have become portable and pervasive automatic identification devices that are widely used in wireless identity identification, Internet of things, medical, security, and public transportation(*1*). These RFID tags can be categorized into three types based on their working frequency, namely, low-frequency (LF) RFID tags operating at 125-135 kHz, high-frequency (HF) tags at 13.56 MHz, and ultra-high-frequency (UHF) tags at over 860 MHz.

The most high-frequency demanding devices in the RFID tags are rectifying diodes, these devices can convert the received alternating current (a.c.) signal into quasi-direct current (d.c.) to power the logic circuit of the tags(*2,3*). In the last two decades, organic materials have received much attention due to their flexible mechanical properties and low processing temperatures(*4-6*), and significant efforts have been made to enhance the performance of organic rectifying diodes (*7-16*). For instance, a typical organic rectifier rectified d.c. voltage at 18V (*17*), fully inkjet-printed polymer diode can rectify at 13.56 MHz (*18*), fully integrated rectifiers able to convert alternating current with a frequency of up to 10 MHz (*19*), and rather thick two-dimensional layered semiconducting crystals rectifying diodes (*20*). The rectifying diode's highest working frequency ($f_{max}$) can be determined by $f_{max} \propto \mu/L^2$, where μ and L are the carrier mobility and thickness of the semiconductor layer, respectively(*17*). The intrinsic low charge carrier mobility largely limits their operation in high frequencies such as in rectifying process.

To overcome this limitation, metal halide perovskites (MHPs) can be proposed as an alternative to organic materials because they possess two orders of magnitude higher mobility(*21,22*), indicating that MHP-based diodes can operate at even higher frequencies. Additionally, MHPs also offer advantages in terms of mechanical properties and low processing temperatures(*23-27*), which have been demonstrated in solar cells(*28-32*), photodetectors(*33-38*) and light-emitting diodes(*39-41*). Nonetheless, MHPs are ionic crystals, the ions like iodide ions inside the matetials may move back and forth and strike the lattice of perovskite repeatedly with the alternating of the electric field when rectifying. Therefore, it is reasonable to assume that the ion-migration may damage the crystal structure and deteriorate the rectifying performance of MHP-based diodes.

In this work, we report rectifying diodes based on MHPs that exhibit high performance when rectifying an incoming sinusoidal signal at 13.56 MHz. We employed a triangular wave detection method to measure the intensity of ion migration at different frequencies. Surprisingly, we found that the ion-migration issue did not occur at such a high-frequency. Our findings suggest that ion-migration will not prevent MHPs from high-frequency usage. Instead, the high carrier mobility of MHPs may facilitate the development of MHP-based RF devices.

**Results and discussions**

In general, the majority carriers in perovskite are assumed uniformly distributed, rectifying diode's highest working frequency $f_{max}$ set by the transit time $t_T$ in an MHP thin film,

$$f_{max} = \frac{1}{t_T} \quad (1)$$

As for the transit time $t_T$,

$$t_T = \frac{L}{v} \quad (2)$$

where $L$ is the thickness of MHPs films, $v$ is the minority carriers' average drifting velocity. The minority carriers' average drifting velocity $v$ can be expressed as:

$$v = \frac{\mu}{E} = \frac{\mu L}{U - U_{d.c.}} \quad (3)$$

where $\mu$ is the mobility of the minority carriers, $E$ and $U$ are the applied electric field and voltage, and $U_{d.c.}$ is the rectified d.c..

Out of equations (1)–(3) we get,

$$f_{max} = \frac{\mu(U - U_{d.c.})}{L^2} \quad (4)$$

From the above, the maximum operating frequency ($f_{max}$) of a diode-based rectifier was determined by $f_{max} \propto \mu/L^2$, indicating that thinner devices can operate at higher frequencies. Therefore, we fabricated thin-film perovskite diodes with three different structures shown in Figure 1a. The perovskite layers in all structures are as thin as 210 nm. We analyzed the J-V characteristic curves of the samples, as shown in Figure 1b and c. Our results indicate that the sample without a hole transport layer (HTL) has approximately two times higher maximum current density than the samples with Poly[bis(4-phenyl)(2,4,6-trimethylphenyl)amine]synonym (PTAA) and even four times higher than the sample with poly(3,4-)ethylenedioxythiophene: polystyrenesulfonate (PEDOT:PSS) under the same voltage bias. When the ITO is replaced by Au, the current density is even ten times higher, but the diode quickly broke down as the voltage bias increases. The rectification ratio of the samples without HTL is $1.9 \times 10^3$ at $\pm 1$

V, and their breakdown reverse voltage could reach 8.5 V, which is higher than the 8 V threshold required to power the logic circuit of the RFID tags. These results indicate that the perovskite-based diodes meet the basic working conditions of rectifying diodes in HF RFID tags, thus making them suitable for HF RFID applications.

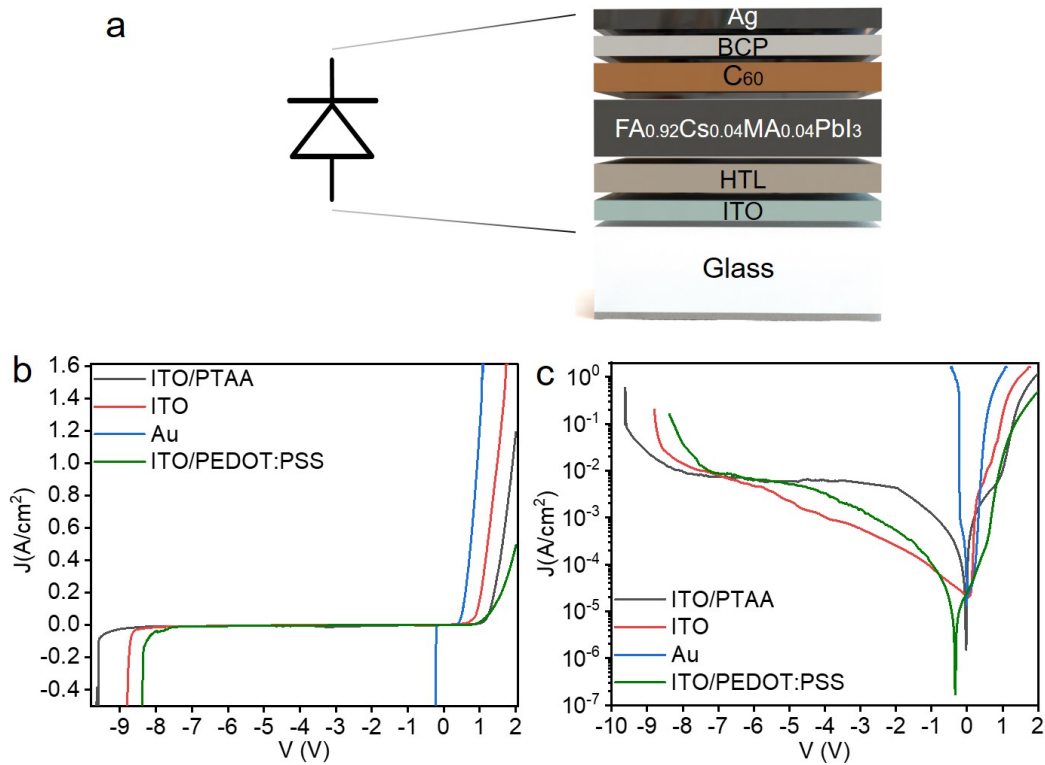

**Figure 1 | Current-voltage curves of metal halide perovskite rectifying diodes. a**, The metal halide perovskite (MHPs) diodes structure schematic. A structure of ITO/ Poly[bis(4-phenyl)(2,4,6-trimethylphenyl)amine]synonym (PTAA) or poly(3,4-)ethylenedioxythiophene: polystyrenesulfonate (PEDOT:PSS)/$FA_{0.92}Cs_{0.04}MA_{0.04}PbI_3$/$C_{60}$/BCP/Ag, despite the indium tin oxide (ITO), their thickness are 40 nm (or 10 nm), 210 nm, 50 nm, 5 nm, 100 nm, respectively. **b**, The diodes' J-V characteristic. **c**, Logarithmic plot of diode's J-V characteristic. The rectification ratio of the diodes without HTL attains $1.9 \times 10^3$ at ± 1 V, and its breakdown reverse voltage is 8.5 V, higher than 8 V threshold required to power the logic circuit of the RFID tags.

The HTLs in perovskite diodes promote extracting holes from the active layer, while they block the transportation of electrons to the electrodes. The energy levels of MHP diodes are shown in Figure2. The energy levels of metal halide perovskite (MHP) diodes are illustrated in Figure 2. Under forward bias, electrons are injected from the Ag electrode (the cathode) and travel through several layers, ultimately reaching the ITO electrode (the anode) under the external electric field. In this process, PTAA and PEDOT:PSS (as shown in Figure 2a and b) act as barriers to electron transport, hindering the flow of electrons to the anode. Conversely, in the absence of an HTL (as shown in Figure 2c), there is no energy barrier for electrons, allowing more electrons to be collected by the anode and resulting in higher current output. When PTAA is used as the HTL in conjunction with perovskite, a small energy barrier is formed, whereas PEDOT:PSS creates a larger energy barrier that impedes the injection of holes. We suggest that a higher barrier results in lower hole injection efficiency in diodes with PEDOT:PSS, leading to a lower current density compared to diodes with PTAA.

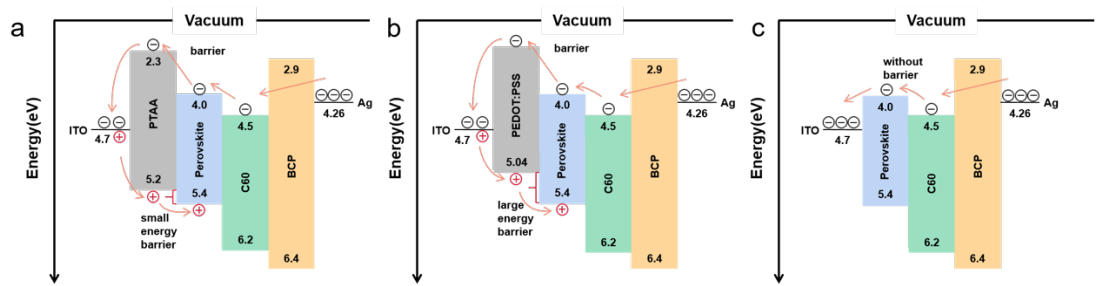

**Figure 2 | Energy levels of different MHP diodes. a**, Energy levels of diodes with PTAA. **b**, Energy levels of diodes with PEDOT:PSS. The energy barriers hinder the transportation of electrons to the anode. **c**, Energy levels of diodes without HTL. The electrons transfer without a barrier between perovskite and ITO. (⊕/⊖: free carriers of holes/electrons)

After designing and fabricating the MHP diodes, we tested their rectifying performance using a high-frequency measurement setup. A typical rectifying bridge, consisting of four rectifying diodes in series with a load resistance ($R_L$) and a load capacitance ($C_L$) in parallel, was used for testing. In order to power the logic circuit of the RFID tags, the rectified direct current(d.c.) signal should provide at least an 8 V supply voltage with minimal voltage drop over the diodes(*42-44*). To measure the performance of the rectifier, we set up a circuit on an electrical testing circuit board, as shown in Figure3. The alternating current (a.c.) input signal was decoupled from the d.c. output signal using a transformer with a primary-to-secondary voltage ratio of 1:1. The transformer, contact needles to the diodes, discrete load resistance $R_L$, and capacitance $C_L$ were arranged on the circuit board. The rectified d.c. signal was monitored using an oscilloscope. All measurements were conducted in air, and the diodes were encapsulated to prevent external interference.

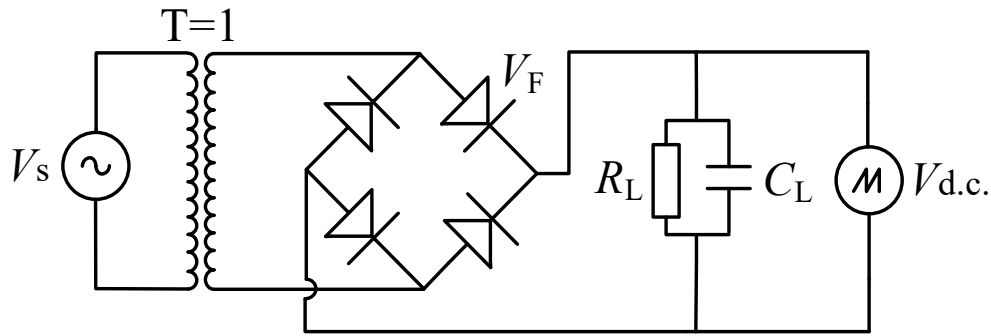

**Figure 3 | High-frequency rectifying measurement setup for MHPs diodes.** The a.c. (alternating current) input signal is decoupled from the d.c. (direct current) output signal by a transformer with a voltage ratio of the primary coil to the secondary coil of 1 to 1. The transformer, the contact needles to the diodes, the discrete load resistance $R_L$ (1MΩ) and capacitance $C_L$ (10µF) are all arranged on a circuit board.

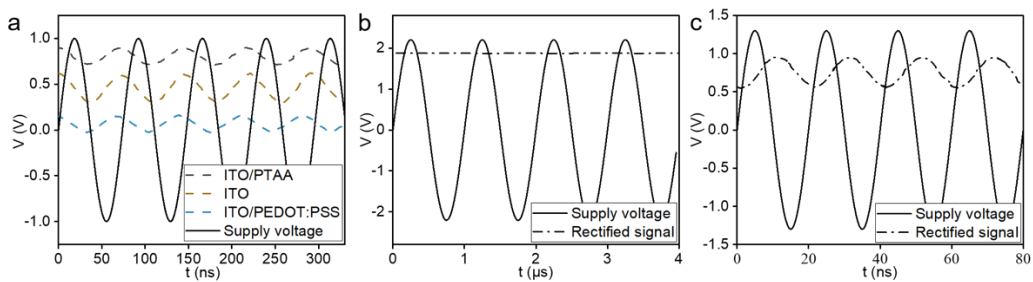

**Figure 4 | High-frequency rectifying measurement of MHP diodes**. **a**, Supply voltage and rectifying signals of diodes with PTAA and PEDOT:PSS and diodes without HTL at 13.56MHz. The rectified signal of diodes with PTAA is higher than diodes without HTL, as well as the diodes with PEDOT:PSS. **b**, Supply voltage and rectifying signals of diodes with PTAA at 1 MHz. **c**, Supply voltage and rectifying signals of

diodes with PTAA at 50 MHz ($A_{diode}$ = 1.3mm × 4.7mm, $R_L$ = 1MΩ, $C_L$ = 10μF).

Figure 4a shows the a.c. supply voltage and rectifying signal of MHP diodes with and without HTL at 13.56 MHz, where the peak-to-peak amplitude of the applied voltage is 2 V. The diodes with PTAA have a rectified signal of around 0.80 V, while the diodes with PEDOT:PSS and those without HTL have rectified signals of 0.10 V and 0.47 V, respectively. At frequencies above 1 MHz, the cable's inductance and capacitance of the discrete $R_L$ and $C_L$ can induce moderate vibrations in the rectified signal. The rectified signal of diodes with PTAA at 1 MHz and 50 MHz is shown in Figures 4b and 4c, respectively. At 1 MHz, the peak-to-peak amplitude of the a.c. supply voltage is 4.5 V, and the resulting d.c. output is 1.87 V without vibration. However, at 50 MHz, the peak-to-peak amplitude of the a.c. supply voltage is reduced to 2.6 V, and the resulting d.c. output is around 0.76 V with strong vibration. The rectified signal decreases with increasing thickness of the perovskite layer, consistent with $f_{max} \propto \mu/L^2$ (Figure S1).

Surprisingly, the rectified signal of diodes with PTAA is higher than that of diodes without HTL, which contradicts the J-V characteristic shown in Figure 1b. This could be attributed to pin-holes on perovskite films. During the spin-coating process of the perovskite precursor on ITO and PTAA substrates, we observed that the precursor is wettable on ITO but shows low wettability on PTAA. On a wettable surface, a non-compact perovskite film with an abundance of pin-holes is formed, even when fabricating a thin film. The current leaks through the pin-holes, resulting in a smaller current flowing into the load resistance. Therefore, diodes without HTL (precursor directly spin-coated on ITO) have a lower rectified signal due to current leakage through these pin-holes. However, on a non-wettable substrate, the film is more compact and without pin-holes, leading to a higher rectified signal.

To develop a pin-hole-free 200 nm perovskite thin-film, we investigated the use of an ultra-thin modified layer of NMABr spin-coated on PTAA, followed by annealing at 100 °C and 150 °C for 5 minutes each. The modified layer significantly improved the wettability of perovskite precursor on PTAA after annealing at 100 °C for 5 mins, while this improvement was limited and even disappeared after annealing at 150 °C for 5 mins. We conducted four experimental groups with varying wettability levels, and the top-view SEM images of the four groups are presented in Figure S2. As expected, the thin perovskite film exhibited more pin-holes with increasing wettability, in agreement with our assumption. Thus, a low wettability of perovskite is necessary to fabricate a thin film without pin-holes.

Additionally, carrier transport speed and lifetime may also affect the rectifying ability of the devices. Faster transport speed corresponds to a higher rectifying frequency, while a long carrier lifetime results in greater rectifying efficiency. To measure the carrier transport time and recombination lifetime of the devices, we performed transient photocurrent (TPC) decay under the short-circuit condition and transient photovoltage (TPV) decay under the open-circuit condition (Figure S3). The charge transport time (measured by TPC) for diodes with PTAA is approximately 251 ns, which is shorter than the 303 ns for diodes without HTL, indicating faster charge transport for diodes with PTAA. Furthermore, the charge recombination lifetime (measured by TPV) for the diodes with PTAA is about 6.26 μs, which is three times longer than the 2.07 μs for the diodes without HTL. Under the open-circuit condition upon illumination, electrons and holes are generated and recombined within the active layer without flowing to the external circuit. When the illumination stops, the separated holes and electrons undergo recombination in the dark, resulting in photovoltage decay. The TPV results indicate that the interface between perovskite and PTAA has fewer defects and traps than the interface between perovskite and ITO. Therefore, charge carriers are less likely to be captured by defects and traps, leading to a faster response at high-frequency.

Additionally, we measured the effect of ion migration using a setup depicted in Figure 5b. In both the simulation and measurement, a triangular wave with a peak-to-peak amplitude of 2 V (Figure 5c) was used as the supply voltage. Given that the iodide ions are confined to the perovskite layer, we simplified the simulation model by neglecting the transport layers. Initially, we allowed the iodide ions to move under the influence of the internal potential electric field without any external voltage applied until they were stably distributed. We then applied a triangular wave at 5, 10, 20, and 100 Hz to the diode. When the supply voltage reached points A, B, C, D, and E in Figure 5c, we simulated the ionic distributions and presented them in Figure 5d, e, f, and g, respectively. As frequency increases, we observed weaker motion of the iodide ions until they barely moved at all. Moreover, we analyzed the actual rectifying signal measured by the bridge rectifier at frequencies of 5, 10, 20, and 100 Hz, as shown in Figure 5h, i, j, and k, respectively. We fit the rise and fall fractions of the signal as lines and recorded their slopes, as well as the rectifying wave's minimum absolute value (Vmin), in Table 1. Slope1 and |Slope2| represent the slope values of the rise and fall fractions, respectively. During the rise fraction of the supply voltage, the ions moved in the direction opposing the applied electric field, accumulating on one side and forming an internal electric field to counter the external electric field. Thus, at the start of the fall fraction of the supply voltage, the total applied electric field would be smaller than the external one, resulting in the rectified signal current. This explains why |Slope2| was greater than Slope1. However, if the ion migration is weaker, the applied voltage will be only slightly affected by it, making the value of |Slope2| closer to Slope1. Due to the diode's properties, carriers should not be injected when applying a negative bias, meaning that the negative current is largely due to ion migration. Consequently, Vmin directly reveals the magnitude of the ion migration current.

Table 1 shows that as the frequency of the supply voltage increases, the value of |Slope2|/Slope1 decreases and is almost equal to 1 at 100 Hz. Additionally, Vmin decreases and can be neglected at 100 Hz. These results indicate that ion migration is extremely weak at such low frequencies and can be ignored at higher frequencies, including 13.56 MHz. Surprisingly, the ions inside perovskite do not damage the device as they cannot keep up with the rapid exchange rate of the external alternating electric bias, and their high carrier mobility enables them to rectify a.c. signals at very high frequencies.

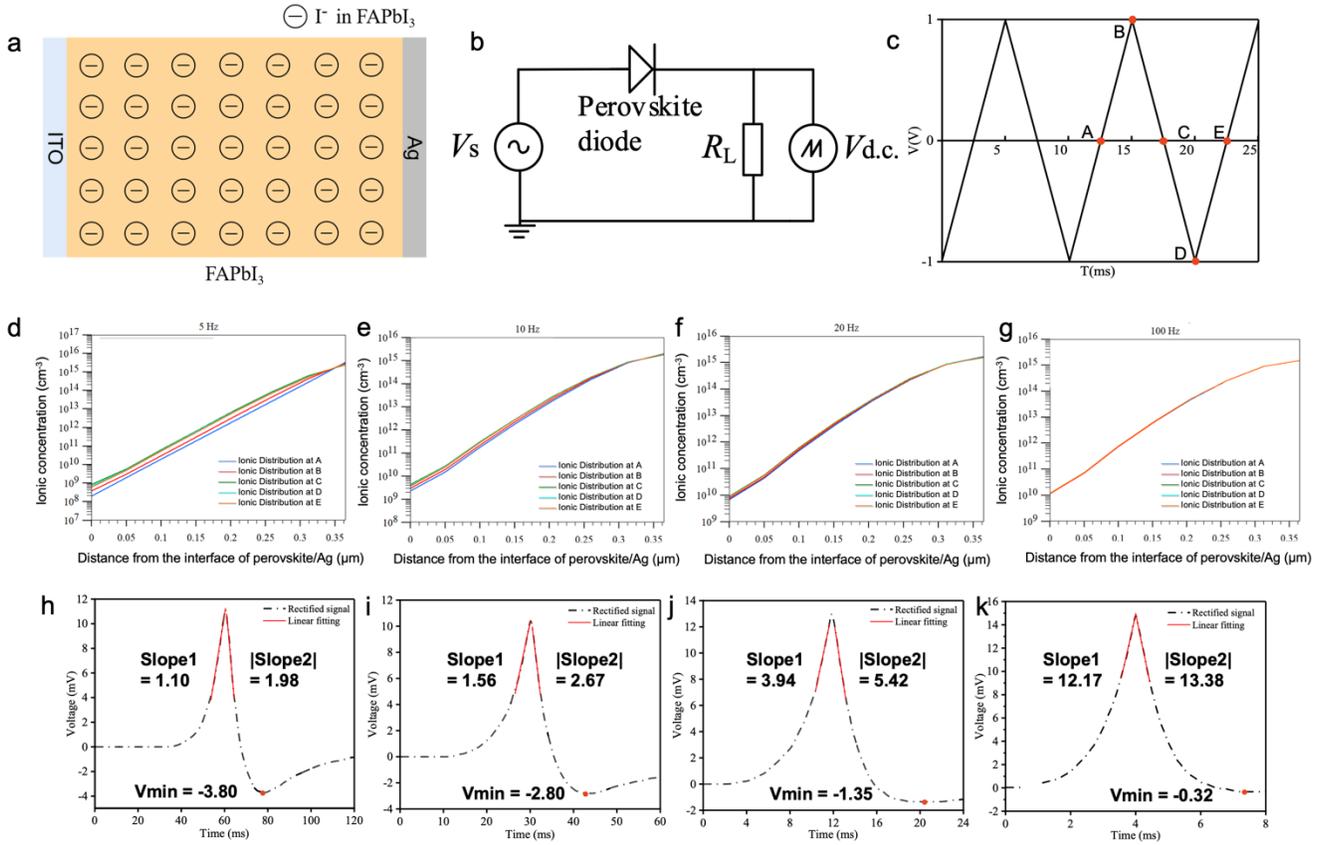

**Figure 5 | Simulation and measurement of ion migration at different frequencies. a**, A simulation model of sandwich structure with iodide ions (I⁻) inside perovskite. **b**, The measurement setup. ($R_L$ = 1MΩ, with the input impedance of oscilloscope set to 50 Ω) **c**, A triangular wave with a peak-to-peak amplitude of 2 V is used as the supply voltage in both simulation and measurement, the ionic distributions when the supply voltage reaches points A, B, C, D, E are simulated. The simulated ionic distribution inside perovskite at **d**, 5 Hz. **e**, 10 Hz. **f**, 20 Hz. **g**, 100 Hz. The measured responses of the triangular wave at **h**, 5 Hz. **i**, 10 Hz. **j**, 20 Hz. **k**, 100 Hz are revealed. The rectifying signal wave's minimum value (Vmin) and |Slope2|/Slope1 represent the degree of ion migration. As the frequency increases, |Slope2|/Slope1 diminishes and is almost equal to 1 at 100 Hz. The wave's minimum absolute value (Vmin) also shows such a trend. The ion migration is very weak at such a low frequency.

**Table 1 Ion migration measurement results in Figure5 (h) – (k).** The rise and fall slope and their ratio,

as well as the minimum voltage were documented.

| Frequency | Slope1 | |Slope2| | |Slope2|/Slope1 | Vmin(mV) |
|---|---|---|---|---|
| 5Hz | 1.10 | 1.98 | 1.80 | -3.80 |
| 10Hz | 1.56 | 2.67 | 1.50 | -2.80 |
| 20Hz | 3.94 | 5.42 | 1.38 | -1.35 |
| 100Hz | 12.23 | 13.38 | 1.09 | -0.32 |

In conclusion, we have successfully demonstrated the development of rectifying diodes based on MHP that are capable of operating at 13.56 MHz. Our experimental results show that the rectification ratio of MHP diodes at ±1 V and breakdown reverse voltage can be up to $1.9 \times 10^3$ and 8.5 V, respectively. Furthermore, we have found that the diodes with PTAA exhibit dynamic properties at high frequencies up to 50 MHz, which can well cover the typical near-field communication range. Importantly, we have investigated the ion migration effect, which is commonly believed to worsen the rectifying capacity of MHP diodes. Interestingly, we have found that this effect does not occur at high frequencies since the ions cannot catch up with the fast exchange speed of the external alternating electric field. Our work provides a promising technological solution for RFIDs and ion-migration and opens up new possibilities for MHP's application at high frequencies.

**Methods**

**Materials.** Formamidinium iodide (FAI, >98.0%), Methylammonium iodide (MAI, >99.0%), Cesium iodide (CsI, >99.0%) and Lead(II) iodide ($PbI_2$, 99.99%) were purchased from TCI. Poly[bis(4-phenyl)(2,4,6-trimethylphenyl) amine]synonym (PTAA, Mn=13514, Mw=27371), Fullerene-$C_{60}$ ($C_{60}$, 99.5%) and Bathocuproine (BCP, >99.8%) and Toluene (99.8%) were purchased from Sigma-Aldrich Company Ltd. Temozolomide (DMSO, 99%), N,N-Dimethylformamide (DMF, 98%), chlorobenzene (CB, 99.8%) and isopropanol (IPA, 99.5%) were purchased from J&K Scientific. Poly(3,4-)ethylenedioxythiophene:polystyrenesulfonate (PEDOT:PSS) was purchased from Heraeus company.

**Fabrication of metal halide perovskite diodes.** The pre-patterned ITO substrates (1.5 × 1.5 cm$^2$) were bought from PsiOTec Ltd and ultrasonically cleaned with soap, deionized water, acetone and IPA in succession for 10 min. The as-cleaned ITO substrates were treated with UV-O$_3$ for 30min and transferred to a N$_2$-filled glovebox. On one of the samples (diodes A), 35 µl PTAA (3.0 mg ml$^{-1}$ in toluene) was spin-coated onto the ITO substrates with a step program at 3000 rpm for 30 s, on the other sample (diodes B), 35 µl PEDOT:PSS (diluted with deionized water in a ratio of 1:2) was spin-coated onto the ITO substrates with a step program at 5000 rpm for 30 s, then the samples were annealed at 100 °C for 10 min and afterward cooled down to room temperature. As for diodes C, there wasn't any hole transport layer spin-coated onto it. The PTAA layer thickness is ~40 nm and the PEDOT:PSS layer thickness is ~10 nm. The perovskite precursor solution was prepared by mixing CsI (14.55 mg) MAI (8.9 mg) FAI (221.5 mg) PbI$_2$ (645.41mg) in 1600 µl DMF and 400 µl DMSO was stirred at room temperature for 60 min and filtered with a 0.22 µm PTFE filter prior to use. To fabricate the perovskite layer on diodes A, B and C, 35 µl precursor solution was spin-coated on the top of PTAA layer (device A), PEDOT:PSS layer (device B) and ITO (device C) by a two-consecutive step program at 1000 rpm for 10 s and 4000 rpm for 30 s, add anti-solvent CB at five-second countdown. The samples were immediately annealed on a hotplate at 100 °C for 10 min. Afterward, 50 nm of C$_{60}$, 5 nm of BCP were thermally evaporated in a separate vacuum chamber (< 5 × 10$^{-4}$ Pa) in sequence. Finally, 100 nm Ag was thermally evaporated in a separate vacuum chamber (< 5 × 10$^{-4}$ Pa) through a metal shadow mask to define an aperture area of 0.062 cm$^2$ by the overlap of the ITO and the Ag.

**Measurements.** The diodes were encapsulated and all measurements were done in air. The d.c. measurements of the diodes were made with a Keithley 2400 sourcemeter. The HF measurements of the rectifier were made with a Tektronix TDS 3054 B with the input impedance set to 1MΩ. The HF supply voltage was provided by a Tektronix AFG 3252. The thickness of the layers was made with Dektak XT surface profiler. Scanning electron microscopy (SEM) images were taken on a Carl Zeiss Ultra 55 electron microscope. The excitation light (515nm) was generated by a femtosecond laser, (Light Conversion Pharos, 1030 nm, <300 fs, 1 MHz).

**Data availability**

The data that support the findings of this study are available from the corresponding author upon reasonable request.

## Acknowledgements

The authors acknowledge funds received from the National Key Research and Development Program of China (2017YFA0207700), the Outstanding Youth Fund of Zhejiang Natural Science Foundation of China (LR18F050001) and the Natural Science Foundation of China (62074136, 61804134, 61874096).

# Supplementary information
# 13.56MHz Rectifying Diodes Based on Metal Halide Perovskite

Peng Jin[1], Xuehui Xu[1], Zeng Chen[2], Chen Xu[1], Tianyu Liu[1], Hanbo Zhu[1], Xinya Chen[1] and Yang (Michael) Yang[1]*

[1]State Key Laboratory of Modern Optical Instrumentation, College of Optical Science and Engineering, Zhejiang University, Hangzhou, Zhejiang, China, 310027.

[2]Center for Chemistry of High-Performance & Novel Materials, Department of Chemistry, Zhejiang University, Hangzhou, Zhejiang, China, 310027.

**Corresponding Author:**
Email:yangyang15@zju.edu.cn[1]

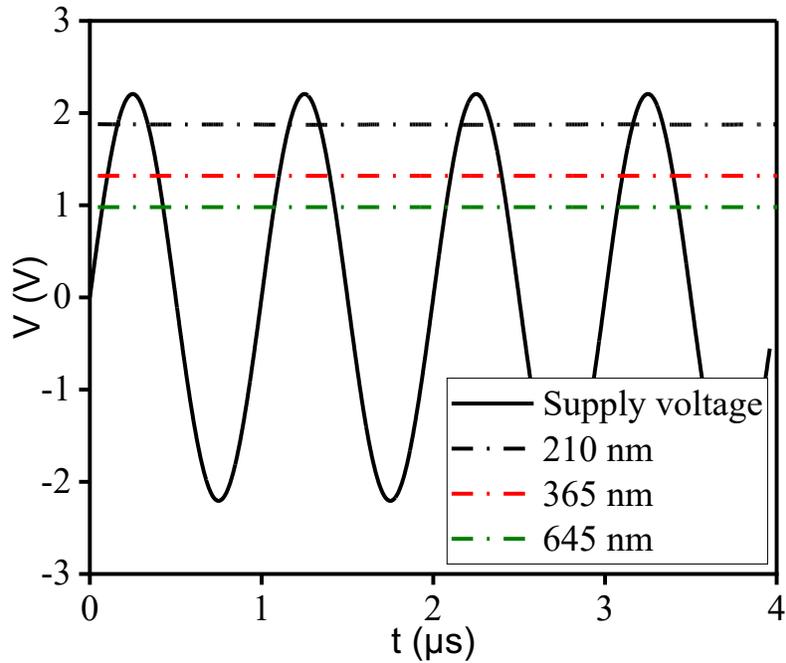

**Figure S1** | Supply voltage and rectifying signals of diodes with different thickness of perovskite layer at 1 MHz. (The structure of the diodes are ITO/PTAA/Perovskite /C60/BCP/Ag)

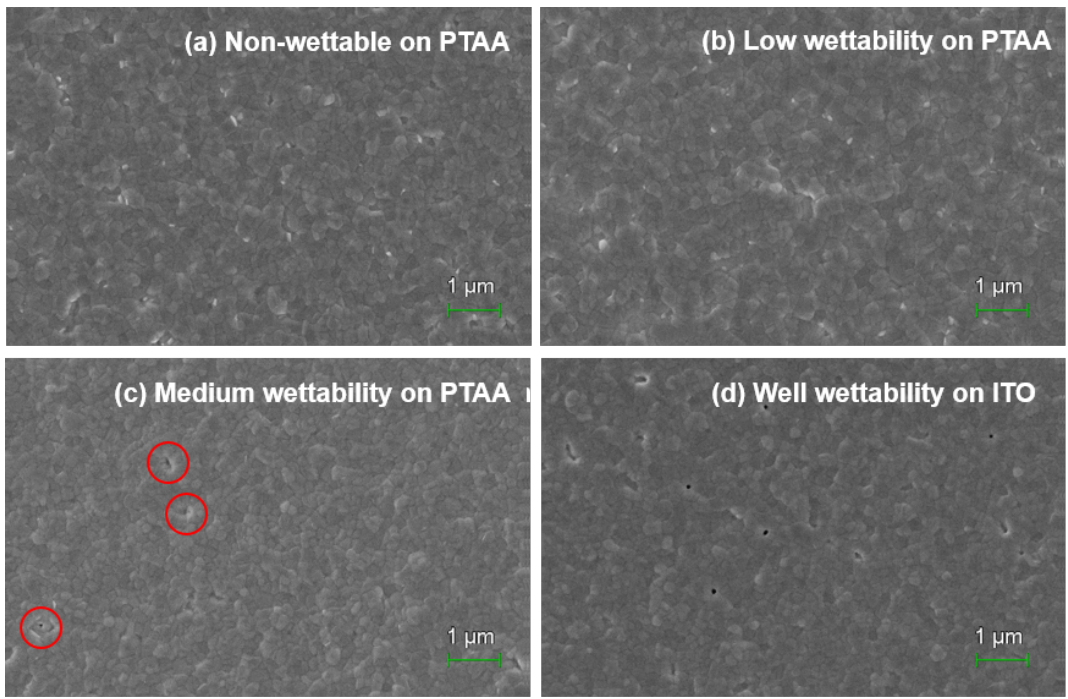

**Figure S2** | Top-view SEM images of perovskite films on PTAA and ITO. **a**, Without the modified layer of NMABr. **b**, Spin-coated a modified layer of NMABr on PTAA, annealed at 150 °C for 5 mins. **c**, Spin-coated a modified layer of NMABr on PTAA, annealed at 100 °C for 5 mins. **d**, Directly

spin-coated perovskite precursor on ITO. The wettability is arranged from low to high in the order of **a** - **d**.

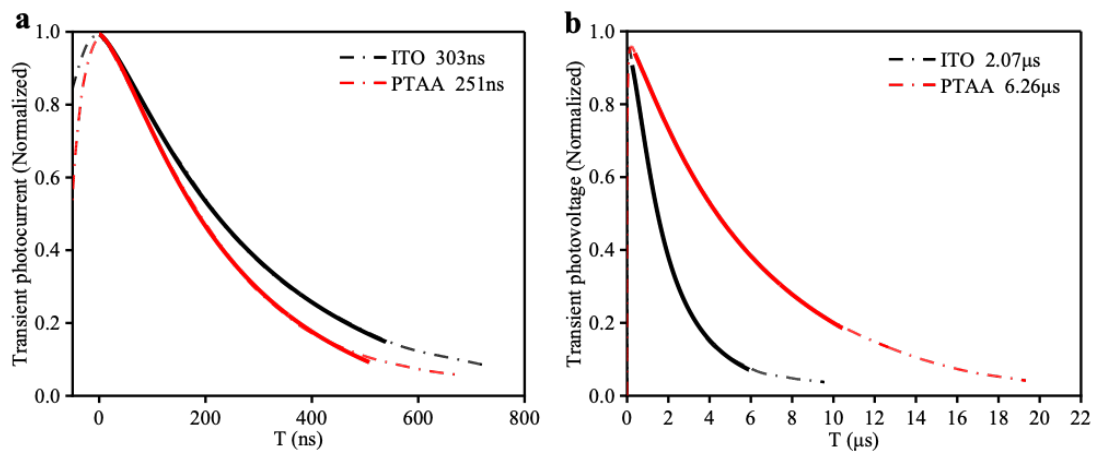

**Figure S3** | Normalized **a**, TPC and **b**, TPV for diodes without HTL and diodes with PTAA.